\documentclass[preprint2]{aastex}
\usepackage{graphicx}
\usepackage{subfigure}
\usepackage{mathrsfs,amssymb}
\usepackage{amsmath}
\usepackage{ulem}

\newcommand{\psec}{\ensuremath{\, {\rm s}^{-1}}}

\newcommand{\km}{\ensuremath{\,{\rm km}}}

\newcommand{\Mpc}{\ensuremath{{\rm Mpc}}}
\newcommand{\K}{\ensuremath{{\rm K}}}

\newcommand{\kHz}{\ensuremath{\, {\rm kHz}}}
\newcommand{\MHz}{\ensuremath{\, {\rm MHz}}}

\begin{document}

\title{Searching for the earliest galaxies in the 21 cm forest}
\author{Yidong Xu$^{1,2,4}$
Andrea Ferrara$^{3}$, Francisco S. Kitaura$^{3}$, and Xuelei
Chen$^{4,5}$}
\affil{$^1$Department of Astronomy, School of Physics, Peking
University, Beijing 100871, China}
\affil{$^2$SISSA/International School for Advanced Studies, Via Beirut 4, 34014 Trieste, Italy}
\affil{$^3$Scuola Normale Superiore, Piazza dei Cavalieri 7, 56126 Pisa, Italy}
\affil{$^4$National Astronomical Observatory of China,
CAS, Beijing 100012, China}
\affil{$^5$Center for High Energy Physics, Peking University,
Beijing 100871, China}

\begin{abstract}
We use a model developed by \citet{XuF10} to compute the 21 cm line
absorption signatures imprinted by star-forming dwarf galaxies (DGs)
and starless minihalos (MHs). The method, based on a statistical
comparison of the equivalent width ($W_\nu$) distribution and flux
correlation function, allows us to derive a simple selection
criteria for candidate DGs at very high ($z\ge 8$) redshift. We find
that $\approx 18\%$ of the total number of DGs along a line of sight
to a target radio source (GRB or quasar) can be identified by the
condition $W_\nu < 0$; these objects correspond to the high-mass
tail of the DG distribution at high redshift, and are embedded in
large HII regions. The criterion $W_\nu > 0.37\kHz$ instead selects
$\approx 11\%$ of MHs. Selected candidate DGs could later be
re-observed in the near-IR by the JWST with high efficiency, thus
providing a direct probe of the most likely reionization sources.\\
\end{abstract}


\section{INTRODUCTION}
The properties of the earliest galaxies, such as their star
formation histories, masses, production of ionizing photons and
their escape fraction, are crucial in understanding the reionization
process, during which the previously neutral intergalactic medium
(IGM) becomes totally ionized. Thanks to the availability of large
ground-based and space telescopes, and improvements in the searching
technique for Lyman Break Galaxies and Lyman-$\alpha$ Emitters, we
are now tracing galaxy formation at progressively higher redshifts
beyond 6 (see \citealt{Bunker09} for a review). Candidates at
%redshifts as high as $z \approx 8$ are newly reported from the
%analysis of the Hubble Ultra Deep Field (HUDF) \citep{Bouwens09}.
redshifts as high as $z \sim 10$ are newly reported from the
analysis of the Hubble Ultra Deep Field (HUDF) \citep{Bouwens09}.
However, it is now believed that the galaxies that produced most of
the necessary (re)ionizing photons were dwarfs \citep{CF07} which
are currently beyond our capability of direct detection. The
forthcoming James Webb Space Telescope (JWST) will have the
capabilities to directly detect the reionization sources at the
faint end of the luminosity function. Still, given their faintness,
long integration times will be required; hence, defining target
candidate reionizing sources will be of primary importance to study
them in spectroscopic detail.

Instead of looking at a specific galaxy directly, the redshifted 21
cm transition of HI traces the neutral gas either in the diffuse IGM
or in non-linear structures, comprising the most promising probe of
the reionization process (see e.g. \citealt{FOB06} for a review).
While the 21 cm tomography maps out the three dimensional structure
of the 21 cm emission or absorption by the IGM against the cosmic
microwave background (CMB) (e.g. \citealt{Madau97,Tozzi00}), the 21
cm forest observation detects absorption lines of intervening
structures towards high redshift radio sources showing high
sensitivity to gas temperature \citep{XuC09}. The problem of the 21
cm forest signatures produced by different kinds of structures has
been addressed by several authors. \citet{Carilli02} presented a
detailed study of 21 cm absorption by the mean neutral IGM as well
as filamentary structures based on the simulations of
\citet{Gnedin00}, but their box was too small to account for large
scale structures and was not able to resolve collapsed objects.
Instead, \citet{Furlanetto02} used a simple analytic model to
compute the absorption profiles and abundances of minihalos and
galactic disks. Later on, \citet{Furlanetto06} re-examined four
kinds of 21 cm forest absorption signatures in a broader context,
especially the transmission gaps produced by ionized bubbles.

Recently, \citet{XuF10} developed a more detailed model of the 21 cm
absorption lines of minihalos (i.e. starless galaxies, MHs) and
dwarf galaxies (star-forming galaxies, DGs) during the epoch of
reionization, explored the physical origins of the line profiles,
and generated synthetic spectra of the 21 cm forest on top of both
high-$z$ quasars and gamma ray burst (GRB) afterglows.
Interestingly, they find that: (i) MHs and DGs show very distinct 21
cm line absorption profiles (ii) they contribute differently to the
spectrum due to the mass segregation between the two populations. It
follows that it is in principle possible to build a criterion based
on the 21 cm forest spectrum to efficiently select DGs.

The goal of this work is to derive the different signatures of DGs
and MHs using a 21 cm spectrum of high-$z$ radio sources, and
provide a criterion to pick DGs lines in the spectrum. For these
candidates, precise redshift information  will be available;
moreover, given the angular position of the background source, the
21 cm forest observation provides an excellent tool for locating the
high-$z$ DGs. The great advantage of using high-$z$ GRBs as
background radio sources is that the follow-up IR JWST observations
after the afterglow has faded away will not be hampered by the
presence of a very luminous source (as in the case of a background
quasar) in the field\footnote{Throughout this paper, we adopt the
cosmological parameters from WMAP5 measurements combined with SN and
BAO data: $\Omega_b = 0.0462$, $\Omega_c = 0.233$, $\Omega_\Lambda =
0.721$, $H_{\rm 0} = 70.1 \km \psec \Mpc^{-1}$, $\sigma_{\rm 8} =
0.817$, and $n_{\rm s} = 0.96$ \citep{WMAP5}}.

\section{METHOD}\label{method}
Here we briefly summarize the main features of the model used in
this work, but refer the interested reader to \citet{XuF10} for a
more comprehensive description. We use the Sheth-Tormen halo mass
function \citep{ST99} to model the halo number density at high
redshift in the mass range  $10^{5-10}M_\odot$, which covers the
minimum mass allowed to collapse \citep{Abel00,ON07} and most of the
galaxies that are responsible for reionization \citep{CF07}. The
dark matter halos have an NFW density profile within the virial
radii $r_{\rm vir}$ \citep{NFW97}, with a concentration parameter
fitted to high-redshift simulation results by \cite{Gao05}; the dark
matter density and velocity structure outside $r_{\rm vir}$ are
described by an ``Infall model''\footnote{Public code available at
http://wise-obs.tau.ac.il/$\tilde{\;\:}$barkana/codes.html}
\citep{Barkana04}. The gas inside the $r_{\rm vir}$ is assumed to be
in hydrostatic equilibrium at temperature $T_{\rm vir}$ in the dark
matter potential, while the gas outside follows the dark matter
overdensity and velocity field.

Once the halo population is fixed, a timescale criterion for star
formation is introduced to determine whether a halo is capable of
hosting star formation. The timescale required for turning on star
formation is modeled as the maximum between the free-fall time
$t_{\rm ff}$ and the H$_2$ cooling time $t_{\rm cool}$
\citep{Tegmark97}, i.e. $t_{\rm SB} = \max \{\, t_{\rm ff},\, t_{\rm
cool}\,\}$. Then star formation activity begins at $t_{\rm s} =
t_{\rm F} + t_{\rm SB}$, where $t_{\rm F}$ is the halo formation
time predicted by the standard EPS model \citep{LC93}. If $t_{\rm
s}$ is larger than the Hubble time at the halo redshift, we define
the system as a {\it minihalo}, i.e. a starless galaxy. The ionized
fraction in a MH is computed with collisional ionization
equilibrium, which depends on its temperature. The gas within
$r_{\rm vir}$ is at the virial temperature, and in the absence of an
X-ray background the gas outside is adiabatically compressed, so
that the temperature is simply $T_{\rm K} = T_{\rm IGM}
(1+\delta)^{\gamma-1}$, where $\gamma = 5/3$ is the adiabatic index
for atomic hydrogen, and $T_{\rm IGM}$ is the temperature of the
mean-density IGM.
%For $z\lesssim100$, it is well approximated by $T_{\rm IGM} \approx
%368.55\, {\rm K} \times {(1+z)^2}/{(1+134)^2}$ \citep{Shapiro06}.
The Ly$\alpha$ photons inside a MH are produced by recombinations, which
are negligible for most MHs that are almost neutral,
but serve as a dominant coupling source for the most massive
MHs which are partially ionized due to their higher $T_{vir}$.

When the criterion $t_{\rm s} < t_{\rm H}$ is satisfied, star
formation occurs within a Hubble time turning the halo into a {\it
dwarf galaxy}. We use a mass-dependent handy fit of star formation
efficiency provided by \citet{SF09}. Adopting the spectra of high
redshift starburst galaxies provided by
\citet{Schaerer02,Schaerer03}\footnote{http://cdsarc.u-strasbg.fr/cgi-bin/Cat?VI/109}
and an escape fraction of $f_{\rm esc}=0.07$ as favored by the early
reionization model (ERM, \citealt{Gallerani08}), we numerically
follow the expansion of the HII region. The gas temperature inside
the HII region is fixed at $2\times10^4\K$, while the temperature of
gas around the HII region is calculated including the Hubble
expansion, soft X-ray heating and the Compton heating. Although the
soft X-ray heating dominates over the Compton heating, its effect is
weak unless the DG has a higher stellar mass and/or a top-heavy
initial mass function. Besides ionization and heating effects, the
DG metal-free stellar population  produces Ly$\alpha$ photons from
soft X-ray cascading, which could penetrate into the nearby IGM and
help to couple the spin temperature to the kinetic temperature of
the gas. Finally, we account for the Ly$\alpha$ background produced
by the collective contribution of DGs.

With the detailed modeling of properties of both MHs and dwarf
galaxies, and an associated Ly$\alpha$ background, we compute the 21
cm absorption lines of these non-linear structures.
%The diffuse
%IGM beyond regions affected by the gravity of non-linear
%structures expands uniformly with the Hubble flow;  then its
%optical depth is \citep{Field59,Madau97,FOB06}
%\begin{equation}
%\tau_{\rm IGM} \,=\, \frac{3\,h_{\rm P}\,c^3 A_{10}}{32\, \pi\,
%k_{\rm B}\,\nu_{10}^2}\, \frac{n_{\rm HI}(z)}{T_{\rm S}\,H(z)},
%\end{equation}
%where $h_{\rm P}$ is the Planck constant, $c$ is the speed of light,
%$A_{10} = 2.85 \times 10^{-15} \psec$ is the Einstein coefficient
%for the spontaneous decay of the 21 cm transition, and $\nu_{10} =
%1420.4 \MHz$ is the rest frame frequency of the 21 cm transition.
%Here $n_{\rm HI}$, $T_{\rm S}$, and $H$ are the neutral hydrogen
%number density, spin temperature, and the Hubble parameter,
%respectively, and a completely neutral IGM is assumed in the early
%stages of reionization. The neutral IGM creates a
The diffuse IGM creates a global decrement in the spectra of
high-$z$ radio sources, on top of which MHs and DGs produce deep and
narrow absorption lines. The 21 cm optical depth of an isolated
object is \citep{Field59,Madau97,Furlanetto02}:
\begin{eqnarray}\label{Eq.tau}
\tau (\nu) &=& \frac{3\,h_{\rm P}\,c^3 A_{10}}{32\, \pi^{3/2}
k_{\rm B}}\, \frac{1}{\nu^2} \nonumber  \\
&\times& \int_{-\infty}^{+\infty} \frac{n_{\rm HI}(r)} {b(r)T_{\rm
S}(r)}\, \exp\left[\,-\, \frac{(u(\nu)-\bar
v(r))^2}{b^2(r)}\,\right] dx,
\end{eqnarray}
where $A_{10} = 2.85 \times 10^{-15} \psec$ is the Einstein
coefficient for the spontaneous decay of the 21 cm transition,
$n_{\rm HI}$ is the neutral hydrogen number density, $T_{\rm S}$ is
the spin temperature, and $b(r)$ is the Doppler parameter of the
gas, $b(r) = \sqrt{\,2\,k_{\rm B}T_K(r)/m_{\rm H}}$. Here $u(\nu)$
is the frequency difference from the line center in terms of
velocity, $u(\nu) \equiv c\, (\nu-\nu_{10})/\nu_{10}$, where
$\nu_{10} =1420.4 \MHz$ is the rest frame frequency of the 21 cm
line, and $\bar v(r)$ is bulk velocity of gas projected onto the
line of sight at radius $r$. Inside of the virial radius, the gas is
thermalized, and $\bar v(r) = 0$, while the gas outside the virial
radius has a bulk velocity contributed from both infall and Hubble
flow, which is predicted by the ``Infall Model.'' The coordinate $x$
is related to the radius $r$ by $r^2 = (\alpha\, r_{\rm vir})^2 +
x^2$, where $\alpha$ is the impact parameter of the penetrating line
of sight in units of $r_{\rm vir}$.

\begin{figure*}
\begin{tabular}{cc}
\includegraphics[totalheight=6cm,width=8cm]{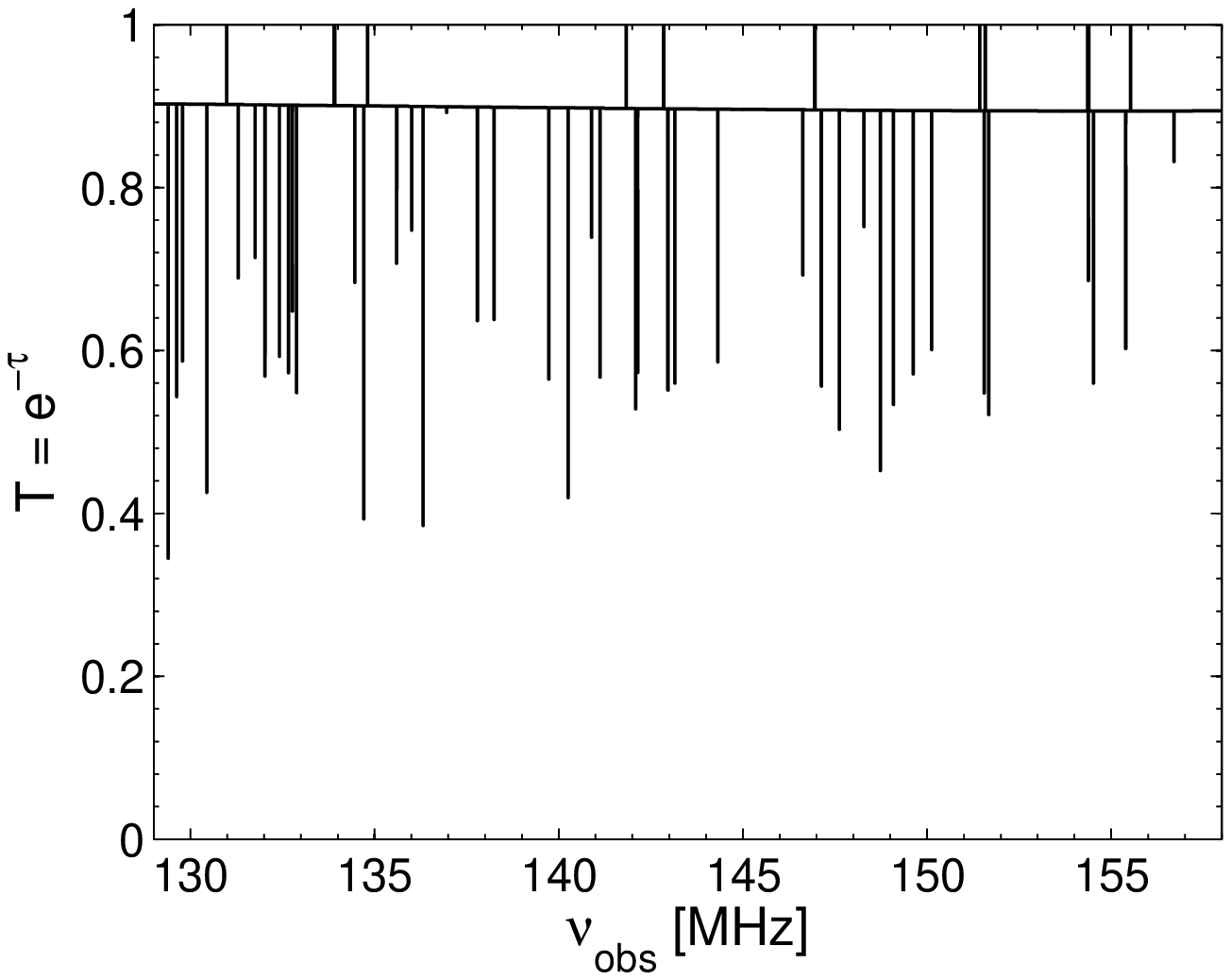}
&\includegraphics[totalheight=6cm,width=8cm]{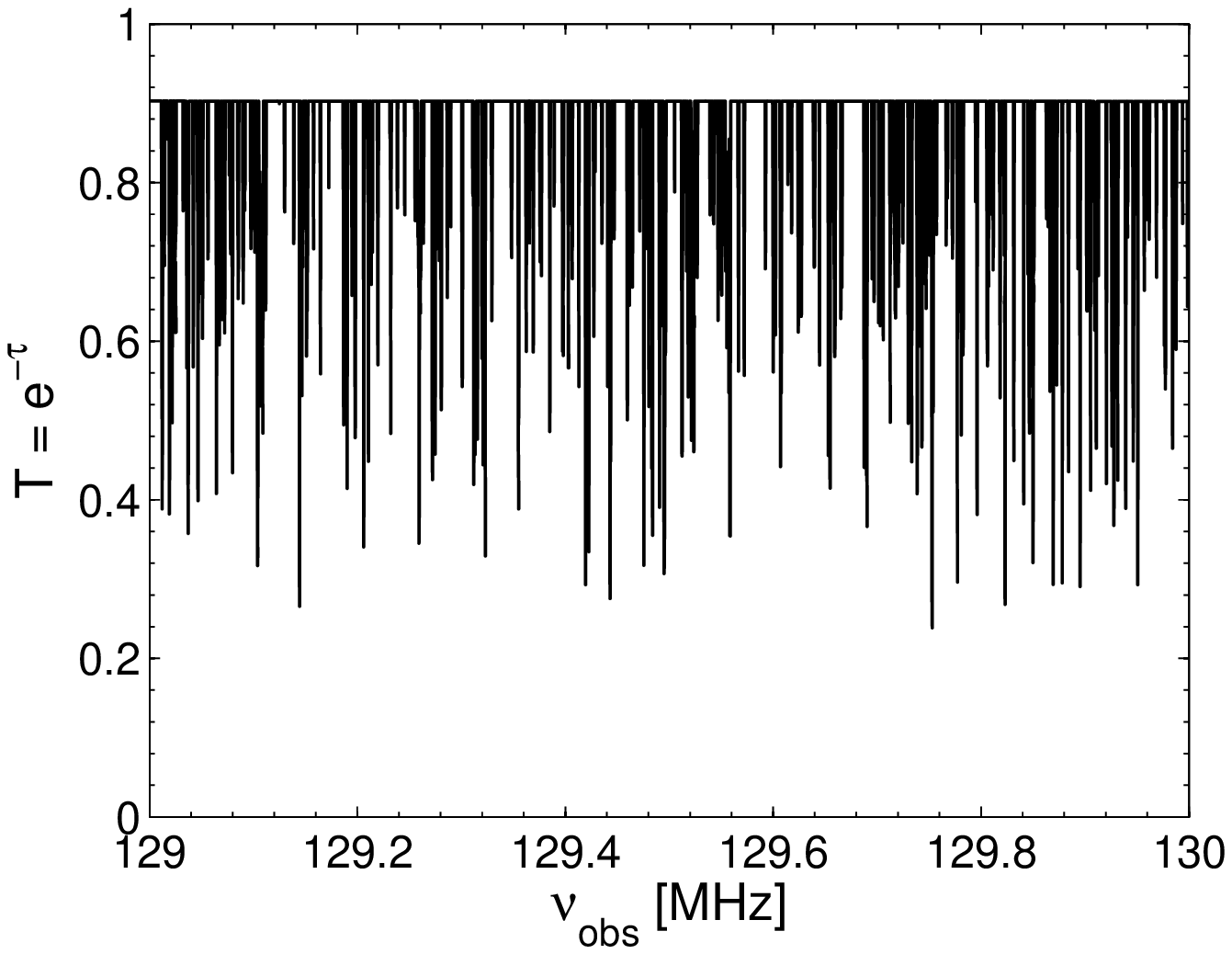}
\end{tabular}
\caption{Relative transmission along a line of sight. {\it Left:}
the spectrum with absorptions caused by DGs alone from 129$\MHz$ to
158$\MHz$ corresponding to $z = 10.01$ -- $7.99$. {\it Right:} the
spectrum with absorptions caused by MHs alone from 129$\MHz$ to
130$\MHz$ corresponding to $z = 10.01$ -- $9.93$.}
\label{Fig.spectrum}
\end{figure*}

The spin temperature of neutral hydrogen is defined by the relative
occupation numbers of the two hyperfine structure levels, and it is
determined by \citep{Field58,FOB06}:
\begin{equation}
T_{\rm S}^{-1} \,=\, \frac{T_\gamma^{-1} + x_c\,T_{\rm K}^{-1} +
x_\alpha\, T_{\rm C}^{-1}} {1+x_c+x_\alpha},
\end{equation}
where $T_\gamma = 2.726(1+z)\,\K$ is the CMB temperature at redshift
$z$, $T_{\rm K}$ is the gas kinetic temperature, and $T_{\rm C}$ is
the effective color temperature of the UV radiation. In most cases,
$T_{\rm C}=T_{\rm K}$ due to the frequent Ly$\alpha$ scattering
\citep{FOB06}. The collisional coupling is described by the
coefficient $x_c$, and $x_\alpha$ is the coupling coefficient of the
Ly$\alpha$ pumping effect known as the Wouthuysen-Field coupling
\citep{Wouthuysen52,Field58}. The main contributions to $x_c$ are
H-H collisions and H-$e^-$ collisions, which can be written as
%\begin{equation}
%x_c \,=\, x_c^{\rm eH} + x_c^{\rm H\!H} \,=\, \frac{n_e\,
%\kappa_{10}^{\rm eH}}{A_{10}}\, \frac{T_\star}{T_\gamma} +
%\frac{n_{\rm HI}\, \kappa_{10}^{\rm H\!H}}{A_{10}}\,
%\frac{T_\star}{T_\gamma},
%\end{equation}
$x_c \,=\, x_c^{\rm eH} + x_c^{\rm H\!H} \,=\, (n_e\,
\kappa_{10}^{\rm eH}/A_{10})\, (T_\star/T_\gamma) + (n_{\rm HI}\,
\kappa_{10}^{\rm H\!H}/A_{10})\, (T_\star/T_\gamma)$, where $T_\star
= 0.0682\, \K$ is the equivalent temperature of the energy splitting
of the 21 cm transition, and $\kappa_{10}^{\rm eH}$ and
$\kappa_{10}^{\rm H\!H}$ are the de-excitation rate coefficients in
collisions with free electrons and hydrogen atoms, respectively.
%These two coefficients at different temperatures are tabulated in \citet{FOB06}.
The coupling coefficient $x_\alpha$ is proportional to the total
scattering rate between Ly$\alpha$ photons and hydrogen atoms,
$x_\alpha \,=\, (4\,P_\alpha/27\,A_{10})\, (T_\star/T_\gamma)$,
%\begin{equation}
%x_\alpha \,=\, \frac{4\,P_\alpha}{27\,A_{10}}\,
%\frac{T_\star}{T_\gamma},
%\end{equation}
where the scattering rate $P_\alpha$ is given by $P_\alpha \,=\,
c\,\sigma_\alpha\, n_\alpha^{\rm tot}/\Delta\nu_D \,=\, 4\pi\,
\sigma_\alpha J_\alpha$.
%\begin{equation}
%P_\alpha \,=\, c\,\sigma_\alpha\, \frac{n_\alpha^{\rm
%tot}}{\Delta\nu_D} \,=\, 4\pi\, \sigma_\alpha J_\alpha.
%\end{equation}
Here $\sigma_\alpha \equiv {\displaystyle \frac{\pi e^2}{m_e c}
f_\alpha}$ where $f_\alpha=0.4162$ is the oscillator strength of the
Ly$\alpha$ transition, $n_\alpha^{\rm tot}$ is the total number
density of Ly$\alpha$ photons, $J_\alpha$ is the number intensity of
the Ly$\alpha$ photons, and $\Delta\nu_D = (b/c)\,\nu_\alpha$ is the
Doppler width with $b$ being the Doppler parameter and $\nu_\alpha$
being the Ly$\alpha$ frequency.

\section{RESULTS}\label{results}

\subsection{The spectra of dwarfs and minihalos}

With the halo number density predicted by the Sheth-Tormen mass
function and the cross-sectional area of halos determined by the
mean halo separation, we derive the number density of the absorption
lines. Applying our star formation criterion to each intersected
halo with Monte-Carlo sampled mass and formation-redshift, we
compute the absorption lines of MHs in collisional ionization
equilibrium or DGs photoionized by central stars, and generate a
synthetic spectrum along a line of sight. The entire spectrum with
absorptions of both MHs and DGs is shown in \citet{XuF10} against a
quasar or a GRB afterglow. In order to illustrate the differences
between the spectrum caused by MHs and that by DGs, and disregarding
the background source properties, we plot the relative transmission
$T=\exp(-\,\tau)$ of a spectrum with absorptions caused by dwarf
galaxies alone (DG-spectrum) in the left panel of
Fig.\ref{Fig.spectrum} and that with absorptions caused by MHs alone
(MH-spectrum) in the right panel, respectively. Note that the ranges
of observed frequency are different in the two panels. The
absorption lines are very narrow and closely spaced in the
MH-spectrum which resembles a 21 cm forest, while the 21 cm lines on
the DG-spectrum are much rarer. A clear signature unique to the
DG-spectrum is that there are some dwarf galaxies with sufficiently
large HII regions that give rise to leaks (i.e. negative absorption
lines with equivalent width $W_\nu < 0$) on the spectrum rather than
absorption lines. Also, we see that absorption lines of MHs are
generally deeper than DGs.

%\begin{figure*}
%\begin{tabular}{ccc}
%\includegraphics[totalheight=4.5cm,width=5.5cm]{autocorrelation_DG_excLeak.eps}
%&\includegraphics[totalheight=4.5cm,width=5.5cm]{autocorrelation_MH.eps}
%&\includegraphics[totalheight=4.5cm,width=5.5cm]{crosscorrelation_MHeDG.eps}
%\end{tabular}
%\caption{The flux correlation functions of the 21 cm forest
%spectrum. {\it Left:} the autocorrelation functions of the
%DG-spectrum. The solid curve includes all the lines while the dashed
%curve excludes the leaks with $W_\nu < 0$. {\it Middle:} the autocorrelation
%function of the MH-spectrum. {\it Right:} the cross-correlation
%function of the DG-spectrum and the MH-spectrum.
%The spectra used for the computation here %of the correlation functions
%are all from 129$\MHz$ to 158$\MHz$ corresponding to $z = 10.01$ --
%$7.99$.} \label{Fig.correlation}
%\end{figure*}

We define $\delta(\nu)$ to be the relative difference between the
flux $f$ at observed frequency $\nu$ and the global flux transmitted
through the homogeneous IGM at the corresponding redshift,
\begin{equation}
\delta(\nu) \,\equiv\, \frac{f(\nu)}{f_{\rm IGM}} - 1.
\end{equation}
Then we compute the {\it flux} correlation functions with the
formula
\begin{equation}
\xi_{ab}(\Delta\nu) \,=\, \frac{1}{N} \sum_{i=1}^{N}\,
\delta_a(\nu_i)\, \delta_b(\nu_i+\Delta\nu),
\end{equation}
where $N$ is the total number of point pairs on the spectrum with a
frequency distance of $\Delta\nu$. The subscript ``ab'' takes the values
``gg'' (``hh'') for the auto-correlation function of the DG (MH)
spectrum, or ``hg'' for the cross-correlation between the MH and
DG spectra.

\begin{figure}
\centering{\resizebox{9cm}{8cm}{\includegraphics{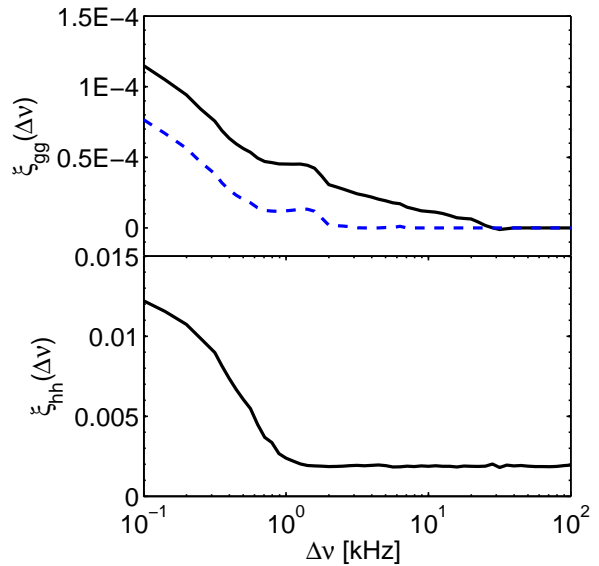}}}
\caption{The flux correlation functions of the 21 cm forest
spectrum. {\it Upper panel:} the auto-correlation functions of the
DG-spectrum. The solid curve includes all the lines while the dashed
curve excludes the leaks with $W_\nu < 0$. {\it Bottom panel:} the
auto-correlation function of the MH-spectrum.
The spectra used for the computation here %of the correlation functions
are all from 129$\MHz$ to 158$\MHz$ corresponding to $z = 10.01$ --
$7.99$.} \label{Fig.correlation}
\end{figure}

We plot these correlation functions with solid curves in
Fig.\ref{Fig.correlation}. From the auto-correlation function of the
DG-spectrum, we see little correlation on frequency separations
larger than $10\kHz$, while little correlation on frequency
separations larger than $1\kHz$ is seen in the MH-spectrum. This is
what we could expect for a randomly generated spectrum with only the
halo mass function. The correlation seen on smaller frequency
separations just comes from the point pairs located within the same
lines. The DGs have a longer correlation length because of their
broader absorption lines or leaks, but the amplitude of the
correlation is very low because they are rare.

In order to illustrate the contribution of those dwarfs with
negative absorption, we calculated the auto-correlation function of
the DG-spectrum with leaks excluded; the result is shown as the
dashed curve in the upper panel of Fig.\ref{Fig.correlation}. The
correlation amplitude is smaller due to the reduced number of
signals, and the correlation length is reduced by more than one
order of magnitude. This shows that the broad signals are caused
primarily by those leaks. In addition, comparing this dashed curve
with the solid curve in the bottom panel, we find that on average,
an absorption line of a DG is even narrower than that of a MH. This
is because some of these DGs produce HII regions larger than $r_{\rm
vir}$, and the absorption lines from the gas outside the virial
radii but inside the HII region, which has the largest infalling
velocities, are erased.

%As for the cross-correlation between MH-spectrum and DG-spectrum, it
%is nearly zero with only stochastic fluctuations: DGs are rare, and
%the transmitted fluxes on a spectrum with only their absorption are
%mostly the flux transmitted through the IGM, and the relative flux
%differences are mostly zero. There is also an interesting albeit
%weak anti-correlation signal in the cross-correlation function.

The cross-correlation function between the MHs and DGs can also be
obtained in the same way. In this work we have not considered the
clustering property of the MHs and DGs arising from large scale
structure, so the flux cross-correlation should be zero, except for
the Poisson fluctuations. Our computation confirms this expectation.
%However, we also find an interesting albeit
%weak anti-correlation signal in the cross-correlation function.

\subsection{Equivalent width distributions}

\begin{figure*}
\begin{tabular}{cc}
\includegraphics[totalheight=6cm,width=8cm]{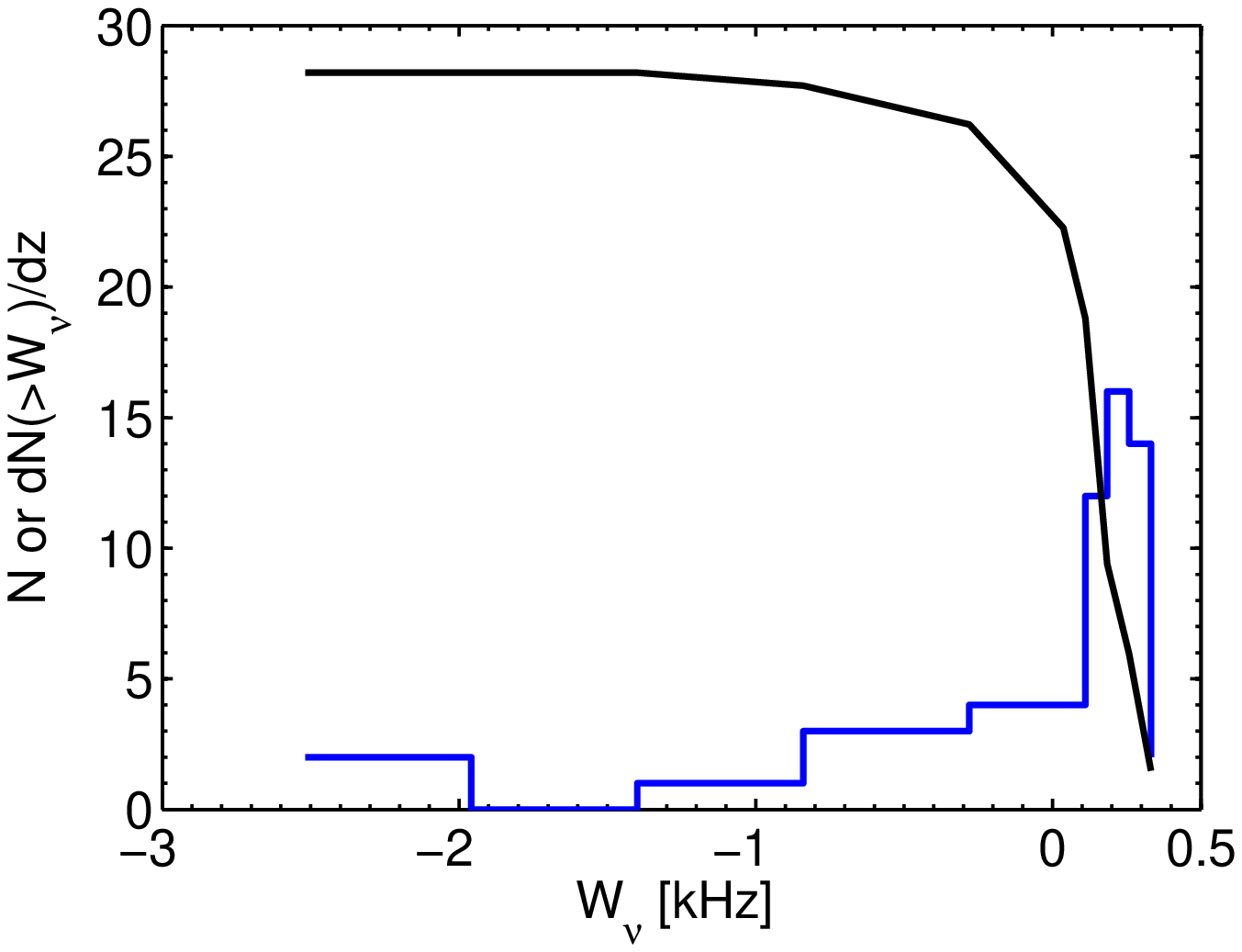}
&\includegraphics[totalheight=6cm,width=8cm]{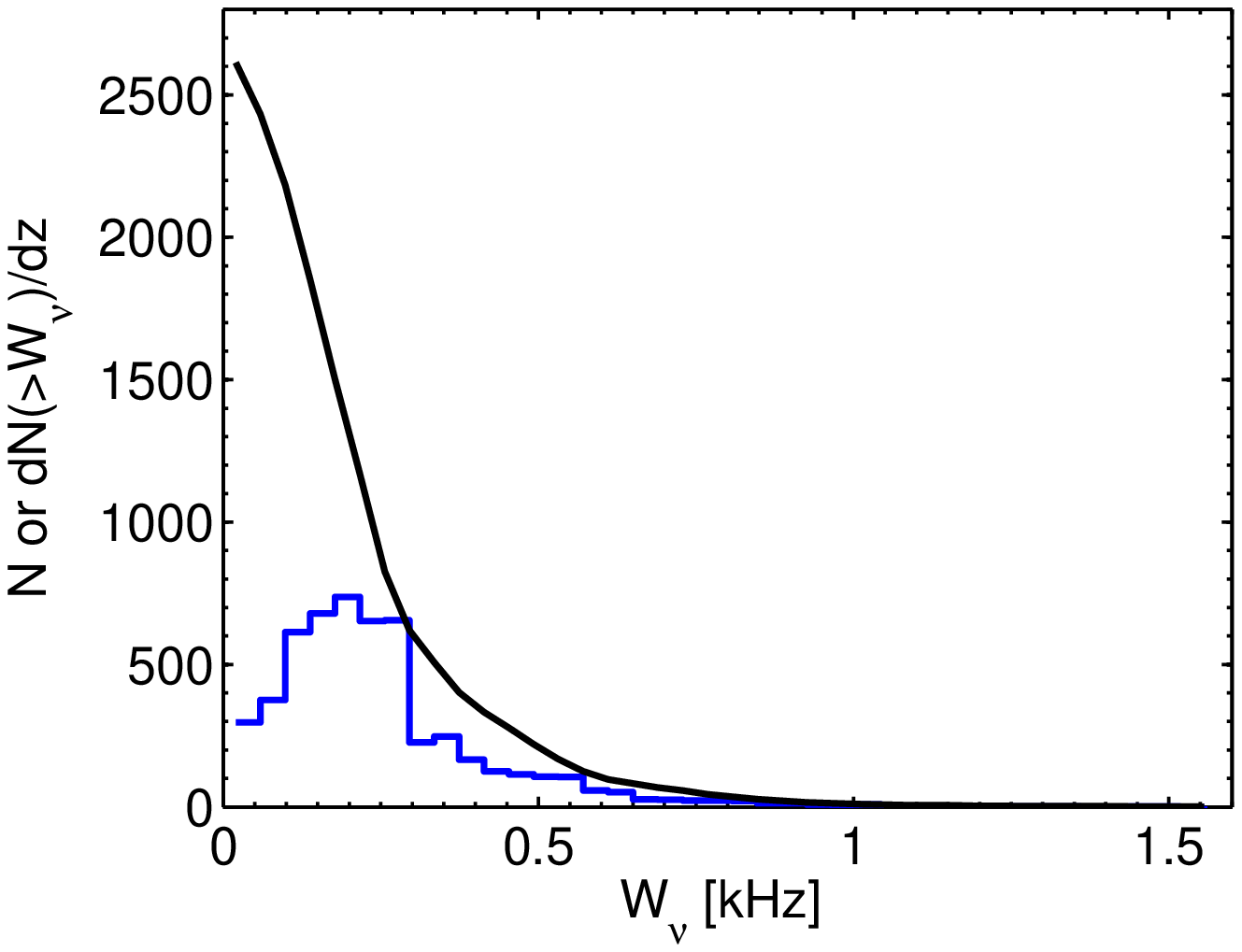}
\end{tabular}
\caption{The differential and cumulative distributions of equivalent
width of the 21 cm absorption lines. {\it Left:} the distribution
for the DG-spectrum. {\it Right:} the distribution for the
MH-spectrum. The spectra used for the computation here %of the equivalent width distributions
are both from 129$\MHz$ to 158$\MHz$ corresponding to $z = 10.01$ --
$7.99$.} \label{Fig.EWdistr}
\end{figure*}

Directly from the spectrum, we could compute the distribution of
equivalent width (EW) of the absorption lines for a specific range
of observed frequency corresponding to a specific redshift. As the
continuum of a background source has a global decrement due to the
absorption of the diffuse IGM, the real signal of non-linear
structures is the extra absorption with respect to the flux
transmitted through the IGM. Therefore, the EW of an absorption
line should be defined as
\begin{eqnarray}
W_\nu &=& \int\, \frac{\displaystyle f_c\, e^{-\,\tau_{\rm IGM}(z)}
\,-\, f_c\, e^{-\,\tau(\nu)}}{\displaystyle f_c\, e^{-\,\tau_{\rm IGM}(z)}}\; d\nu \nonumber \\
&=& \int\, {\displaystyle (1 \,-\, e^{\tau_{\rm IGM}(z) \,-\,
\tau(\nu)})}\; d\nu,
\end{eqnarray}
where $f_c$ is the continuum flux of the background radio source,
and $\tau_{\rm IGM}(z)$ is the optical depth of the diffuse IGM at
redshift $z$. We compute the differential and cumulative
distributions of EW for the DG-spectrum in the left panel and for
the MH-spectrum in the right panel respectively in
Fig.\ref{Fig.EWdistr}. The histograms represent the number
distributions and the solid curves are the cumulative distributions
per redshift interval.

We see that both EW distributions for DGs and MHs
peak in the same region at $W_\nu \sim 0.1$ - $0.3 \kHz$. This
means that most of the dwarfs and MHs have comparable EWs,
and we cannot distinguish them only from their EWs. However, the
distribution curves of their EWs show different shapes. The EW
distribution of DGs has a long tail at the small EW
end, while MHs have a large-EW tail in the distribution
curve.

\subsection{Selection criteria}

In this subsection, we aim at deriving a criterion to distinguish
between the absorption lines caused by DGs and those caused by MHs.

From the computation of EW distributions, we find that only dwarf
galaxies cause negative absorptions and thus have $W_\nu < 0$, while
only MHs are found to have EWs above $0.37\kHz$. Therefore, the
first criterion for candidate DGs could be $W_\nu < 0$, with which
we select 10 dwarfs out of 54 in our synthetic spectrum. They have a
100 percent probability of being caused by DGs. That means, we can
find $18.5\%$ of the total dwarfs along the line of sight, and they
are relatively large dwarfs with large HII regions. In addition,
with the predicted EW distribution, we can estimate the total number
of DGs in the spectrum with the number of selected dwarf galaxies
that have negative absorptions. Similarly, using the second
criterion $W_\nu > 0.37\kHz$, we can select 812 MHs out of a total
of 7108. This is $11.4\%$ of all the MHs, which cannot be
misidentified as DGs.

From the correlation functions shown above, on the other hand, we
see that DGs have a longer correlation length than MHs. In the
absence of halo clustering information, the correlation length
reflects the mean width of the absorption lines, so this is also a
distinctive signature of dwarfs from MHs. However, we have
demonstrated that the correlation of dwarfs at relatively large
frequency distances are exactly caused by those with negative
absorptions. Therefore, the criterion of broad absorption is
degenerate with the negative tail of the EW distribution of the DGs.
Excluding those dwarfs with negative absorptions, the mean width of
an absorption line of a DG (about $0.3\kHz$) is even smaller than
that of an MH (about $0.5\kHz$). Hence, for the lines with $0 <
W_\nu < 0.37\kHz$, if we have infinite resolution, a narrower
absorption line will have a higher probability of being caused by a
dwarf galaxy. This is probably beyond the resolution capabilities of
currently planned instruments. The probability of these absorption
lines being a DG would be $\sim 44/6296 \sim 0.7\%$, with the
complementary probability attributed to MHs.

\section{DISCUSSION}\label{discuss}

Using the model developed by \citet{XuF10}, we have computed 21 cm
absorption line spectra (``forest'') caused by DGs and MHs
separately, their flux correlation functions and EW distributions,
with the aim of distinguishing DGs from MHs in a statistical way.
With the selection criterion of $W_\nu < 0$, we are able to identify
$\sim 18.5\%$ of DGs, and the criterion of $W_\nu > 0.37\kHz$
selects $\sim 11.4\%$ of MHs. As a whole, we can disentangle $\sim
11.5\%$ of all the non-linear objects along a line of sight for
which we can tell whether they are DGs or MHs. In this way, we find
a strong but simple criterion to select candidate DGs to be later
re-observed in the optical or infrared. Using the radio afterglow of
a high redshift GRB as the background, this selection strategy could
be accomplished by LOFAR or SKA. Then, after the GRB fades away, the
follow-up observations can be carried out by JWST, which will be
capable of directly detecting the DGs that are responsible for
reionization.

Cosmic voids can also produce negative absorptions with respect to
the mean absorption by the IGM. Accounting for the density voids
requires the clustering information of large scale structure which
is not included in our computation. However, according to the void
size distribution based on the excursion set model developed by
\citet{SW04}, the characteristic scale of a density void is much
larger than that of a DG HII region. As a result, density voids will
produce ``transmissivity windows'' which are about one order of
magnitude broader than the ``leaks'' produced by HII regions. As the
width of both ``transmissivity windows'' and ``leaks'' exceed the
current spectral resolution, a second criterion of signal width
could be applied to eliminate those voids. Further, it is not
necessary to consider the so-called ``mixing problem'' between the
density voids and the HII regions as \citet{Shang07} did for the
Ly$\alpha$ forest, because the dwarfs are more likely to exist in
filaments out of the voids, and they are not likely to mix with the
density voids.

%Voids: 1. density voids -- $\Delta \nu_V > \Delta \nu_{DG}$. 2.
%density voids vs. galaxy voids (090912). 3. no mixing problem
%between HII-leak and void induces transmissivity window (090912,
%quote Ly$\alpha$ mixing)

While the selection criterion for candidate DGs is reliable, the
total number of predicted DGs and the percentages of identifiable
objects are model-dependent. Specifically, they depend on the star
formation law and efficiency, stellar initial mass function, and
$f_{\rm esc}$. However, the fraction of dwarfs having negative
absorption depends only on the {\it shape} of the EW distribution.
If different star formation models predict similar shapes of the EW
distribution, then our prediction of the fraction of dwarfs
producing leaks is quite reliable and model-independent, and the
total number of DGs along a given line of sight can be safely
inferred from the number of selected leaks. Otherwise, we could
compare the total number of dwarfs inferred from the percentage
argument with the one originally predicted from our star formation
model, and use this result to constrain the model.

To improve on the current selection criteria, the next step is to
include the clustering properties of dark matter halos. With this
ingredient included, the correlation functions will retain
additional information on the distances between the lines. In
principle, knowing the shape of the correlation function, one could
associate to any given line in the spectrum (e.g. by using Bayesian
methods) the statistical probability that it arises from a DG. We
reserve these and other aspects to future work.

\section{ACKNOWLEDGMENTS}
We thank R. Barkana who provided his infall code. This work was
supported in part by a scholarship from the China Scholarship
council, by a research training fellowship from SISSA astrophysics
sector, by the NSFC grants 10373001, 10533010, 10525314, and
10773001, by the CAS grant  KJCX3-SYW-N2, and by the 973 program No.
2007CB8125401.

{}

\end{document}